\documentclass[aps,pra,twocolumn,superscriptaddress,showpacs,floatfix]{revtex4-1}
\usepackage{times,graphicx,psfrag,amsfonts,amsmath,amssymb,color}

\newcommand{\Tr}{{\rm Tr}}
\def\sb{{\scriptscriptstyle B}}

\def\sg{{\scriptscriptstyle N\!G}}
\def\sv{{\scriptscriptstyle V}}

\renewcommand{\prl}{{Phys. Rev. Lett.} }

\newcommand{\ket}[1]{\left\vert #1 \right\rangle}

\begin{document}
\title{Nonlinearity and nonclassicality in a nanomechanical resonator}
\author{Berihu Teklu}\email{berihut@gmail.com}
\affiliation{Dipartimento di Fisica, Universit\`a degli 
Studi di Milano, I-20133 Milano, Italy.}
\affiliation{
Institut Pascal, PHOTON-N2, Clermont Universit\'e, Blaise Pascal
University, CNRS, F-63177 Aubi\`ere Cedex, France.
}
\author{Alessandro Ferraro}\email{a.ferraro@qub.ac.uk} 
\affiliation{School of Mathematics and Physics, Queen's University, Belfast
BT7 1NN, United Kingdom.}
\author{Mauro Paternostro}\email{m.paternostro@qub.ac.uk} 
\affiliation{School of Mathematics and Physics, Queen's University, Belfast
BT7 1NN, United Kingdom.}
\author{Matteo G. A. Paris}\email{matteo.paris@fisica.unimi.it}
\affiliation{Dipartimento di Fisica, Universit\`a degli Studi di Milano, I-20133
Milano, Italy.}
\affiliation{CNISM, UdR Milano Statale, I-20133 Milano, Italy.}
\date{\today}
\begin{abstract}
We address quantitatively the relationship between the nonlinearity of a
mechanical resonator and the nonclassicality of its ground state.  In
particular, we analyze the nonclassical properties of the nonlinear
Duffing oscillator (being driven or not) as a paradigmatic example of a
nonlinear nanomechanical resonator. We first discuss how to quantify 
the nonlinearity of this system and then show that the nonclassicality of 
the ground state, as measured by the volume occupied by the negative part 
of the Wigner function, monotonically increases with the nonlinearity in 
all the working regimes addressed in our study. Our results show 
quantitatively that nonlinearity is a resource to create nonclassical 
states in mechanical systems.
\end{abstract}
\date{\today}
\maketitle
\section{Introduction}\label{s:intro}
Mechanical systems are emerging as very well suited candidates for the
study of quantum behavior at the mesoscopic scale. Such a possibility is
currently being explored in particular in the domain of quantum
opto-/electro-mechanics~\cite{Aspelmeyer}, where ground-breaking
experimental demonstration of quantum control of massive systems
operating under explicitly adverse conditions have been recently
made~\cite{Rogers}. Yet, despite the substantial number of studies
addressing the features of mechanical systems operating at the quantum
level, only partial attention has so far been given to the physics of
nonlinear mechanical devices. Classically, it is  known that many
nonlinear systems exhibit very complex behaviors of potentially
interesting features~\cite{holmes}, and it is currently believed that
such features might be potentially useful in many areas of
investigation, even beyond physics. 
\par
In the context of mesoscopic quantum behaviors, Katz {\it et al.} 
have studied the quantum-to-classical transition in the state of 
nonlinear nanoelectromechanical systems (NEMS) \cite{katz}. Their analysis
considered both an isolated resonator and one open to the effects of an
environment, focusing on quantum signatures and on their disappearance
toward classicality, as the operating temperature of the oscillator was
raised. 
\par
In this paper, we study the link between the enforcement of nonlinearity
in a quantum mechanical oscillator and the manifestation of evidently
nonclassical features in its state. Wondering about such a connection is
indeed significant: the expectation values taken by observables of linear
systems follow the corresponding classical equations of motion, and a
linear dynamics do not let phase-space non-classicality emerge (even at
low temperatures). It is thus important, and relevant for applications, 
to understand which is the
interplay between the nonlinear character of the evolution of a given
bosonic system and the strength of the nonclassical features that we are
able to correspondingly enforce. 
\par
From a quantitative viewpoint, here we will make use of the negativity
of the Wigner function as a phase-space indicator of
nonclassicality~\cite{ana}. This is an established notion of
nonclassicality, with a close relationship with the non-local properties
of the quantum state \cite{coh97,ban98}.  On the other hand, we will
quantify the degree of nonlinearity of a given system by making use of
the measures put forward by some of us in Ref.~\cite{paris14}. In order
to complement the analysis presented in Ref.~\cite{katz}, we focus on
the case of Duffing-type nonlinearity which, besides {being
technologically relevant as inherent in some forms of NEMS
\cite{postma}, has been the focus of a few studies on the quantum
effects that it entails in terms of non-classicality \cite{Duffing_NCl}
and entanglement \cite{Duffing_Ent}.} We show that a direct link exists
between nonlinearity and the nonclassical character of the state of the
oscillator. By focusing explicitly on the ground state of the system, we
show that the phase-space nonclassicality of such state depends almost
linearly on the degree of nonlinearity of the oscillator, thus
suggesting a potential role of the latter feature as a resource for the
achievement of strong quantumness. Our conclusions are valid for a
driven and an undriven Duffing oscillator, thus covering a vast range of
physically relevant situations. We believe that our analysis and results
embody a first interesting step towards the establishment of a rigorous
link between such fundamental features in the dynamics of an oscillator. 
\par
The remainder of the paper is structured as follows. In Sec.~\ref{s:NLN}
we discuss the specific example of nonlinear oscillator, a Duffing
oscillator, considered in this work. We focus on the undriven
configuration of such oscillator and introduce the relevant measures of
nonlinearity and nonclassicality that will be used throughout the paper.
We show that a direct correspondence between degree of nonlinearity and
nonclassicality can be established, pointing at the relevance of former
for the enforcement of the latter in the ground state of an oscillator.
Sec.~\ref{s:DRIVEN} deals with the case of a driven Duffing oscillator,
and reports an analysis similar to the one presented in
Sec.~\ref{s:NLN}. We show that the relation between measures of
nonlinearity and non classicality is maintained in a dynamical situation 
as well, thus highlighting the fundamental nature of the relationship that
we find, which appears to be unrelated to the details of the working
conditions of the oscillator. Finally, Sec.~\ref{conc} closes the paper
with some concluding remarks.
\section{Nonlinearity of a Duffing oscillator: Undriven case} 
\label{s:NLN}
We consider a Duffing oscillator, which is described by the
Hamiltonian~\cite{Carr}  
\begin{equation} 
\label{HamiltonianModel}
 \hat H_{sys}=\frac{1}{2}(\hat p^2+\hat x^2)+
 \frac{1}{4}\varepsilon \hat x^4-\hat x {\cal F}\cos\omega t,
\end{equation}
where $\hat x$ and $\hat p$ are the dimensionless  position- and
momentum-like operators of the oscillator (such that $[\hat x,\hat
p]=i$), $\varepsilon$ is the anharmonicity parameter, ${\cal F}$ is the
amplitude of a possible force that drives the oscillator at frequency
$\omega$. Without loss of generality, throughout the manuscript we will
consider the case of a stiffening nonlinearity with $\varepsilon>0$.
Such model is appropriate to describe the energy of a small-size doubly
clamped mechanical resonator such as a carbon nanotube, or a
nanowire~\cite{postma}. The onset of nonlinear effects in such systems
decreases with decreasing diameter of the device, making either weak
driving forces or thermomechanical noise sufficient to drive the motion
away from the linear approximation. In this Section we will focus on the
undriven case, thereby setting ${\cal F}=0$, deferring the 
treatment of a driven one to the next Section.
\par
\begin{figure}[h!] 
\includegraphics[width=0.49\columnwidth]{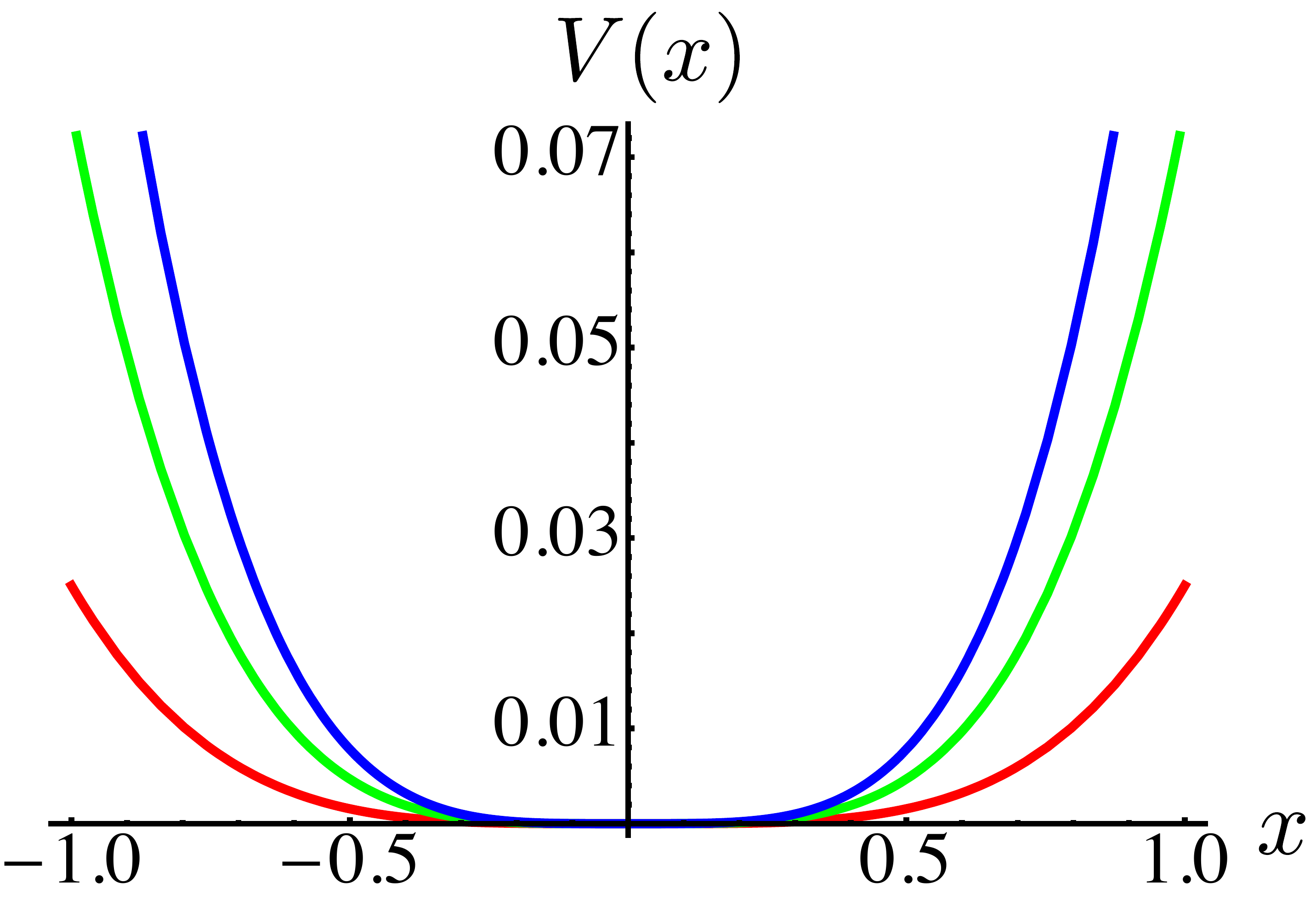}
\includegraphics[width=0.49\columnwidth]{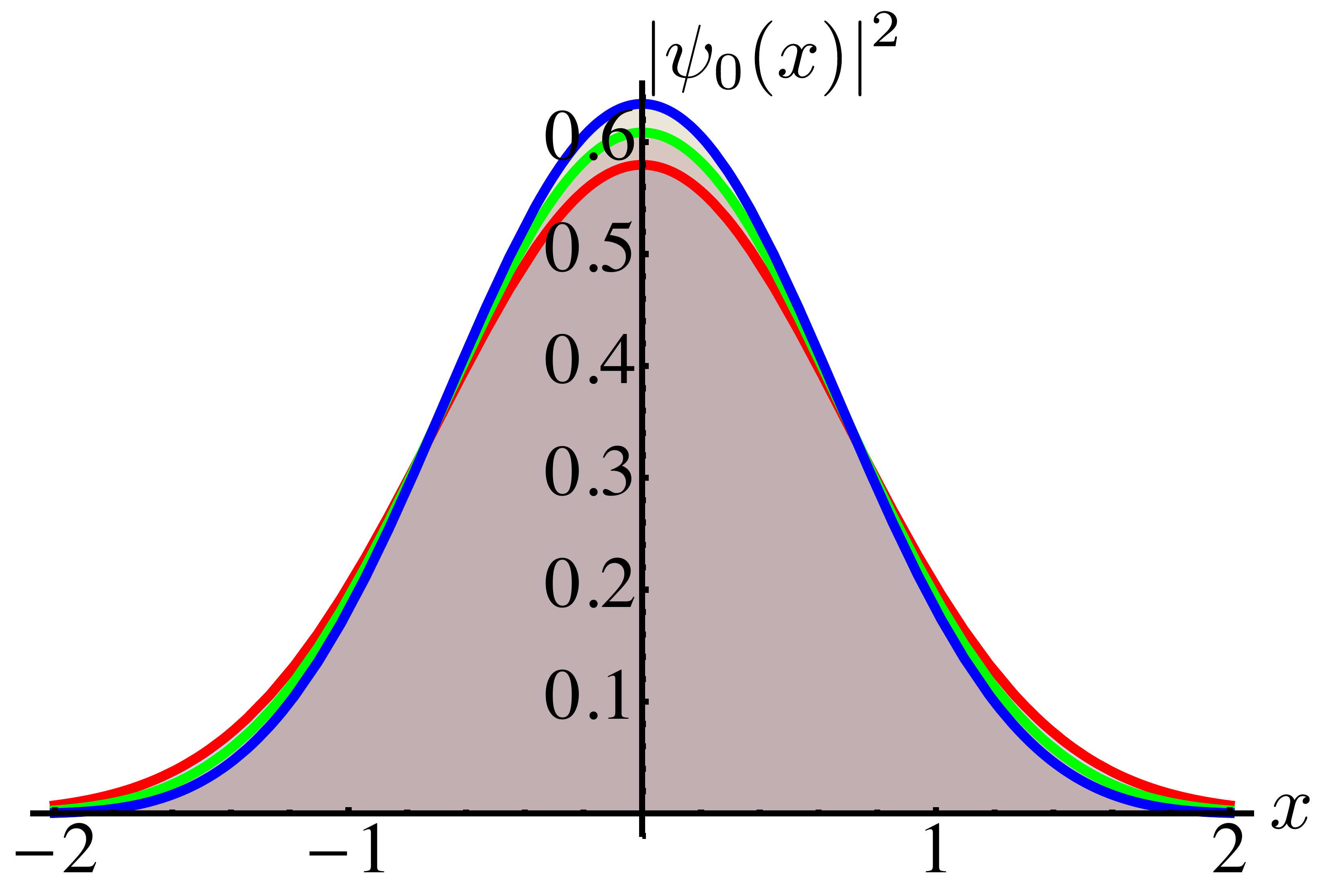}
\caption{(Color online) The Duffing potential. In the left panel we show
$V(x)$ for $\varepsilon=0.2$ (red),  $\varepsilon=0.5$ (green),
$\varepsilon=0.8$ (blue).  In the right panel we show the corresponding
GS probability densities, $|\psi_0|^{2}(x)$.}
\label{f:F1}
\end{figure}

It is instructive to gather an understanding of the form of the
nonlinear potential energy to which the oscillator is subjected. 
In the left panel of
Fig.~\ref{f:F1} we show the function $V(x)=\varepsilon x^4/4$ for
different choices of the anharmonicity parameter, which shows that an
increasing value of $\varepsilon$ results in more pronounced nonlinear
effects at smaller displacements from the equilibrium position $x=0$ of
the oscillator. The effects of the quartic potential on the wave
functions of the oscillator can be evaluated by using
time-independent perturbation theory. In such context, we use the
notation $\hat H_{sys}=\hat H^{(0)}+\hat V(\hat x)$ with
\begin{equation}
\label{modello}
\hat H^{(0)}=\frac12(\hat x^2+\hat p^2),
\qquad\hat V(\hat x)=\frac14\varepsilon\hat x^4.
\end{equation}
We thus evaluate the first-order corrections to the eigenstates $\{\ket{n}\}$ of $\hat H^{(0)}$  as
\begin{equation} 
|\psi_{n}\rangle\approx |n\rangle+\sum_{k\neq n}
\frac{\langle k|\hat V(\hat x)|n\rangle}{n-k}|k\rangle.
\end{equation} 
We will mostly focus on the ground state (GS) of the nonlinear
oscillator, for which we aim at finding an approximate form. While a
fully numerical approach could be used to gather the full form of the GS
of the oscillator, the range of values that $\varepsilon$ can take
experimentally well justifies a perturbative approach.  The perturbing
Hamiltonian only couples $|0\rangle$ to $|2\rangle$ and $|4\rangle$, so
that we have the normalised
approximation of the GS
 \begin{equation} 
|\psi_{0}\rangle={\cal N}\left[|0\rangle-\frac{3\varepsilon}
{8\sqrt{2}}|2\rangle-\frac{\sqrt{3}\varepsilon}{16\sqrt{2}}|4\rangle\right]
\label{eq:GS}
\end{equation} 
with ${\cal N}=1/\sqrt{1+\frac{39 \varepsilon^2}{512}}$. The
corresponding probability densities are plotted on the right panel of
Fig.~\ref{f:F1} for increasing values of $\varepsilon$, showing how the
nonlinear term in the potential tends to localize the wave function of
the oscillator around its equilibrium position. 
\par
\begin{figure}[h!] 
\includegraphics[width=0.85\columnwidth]{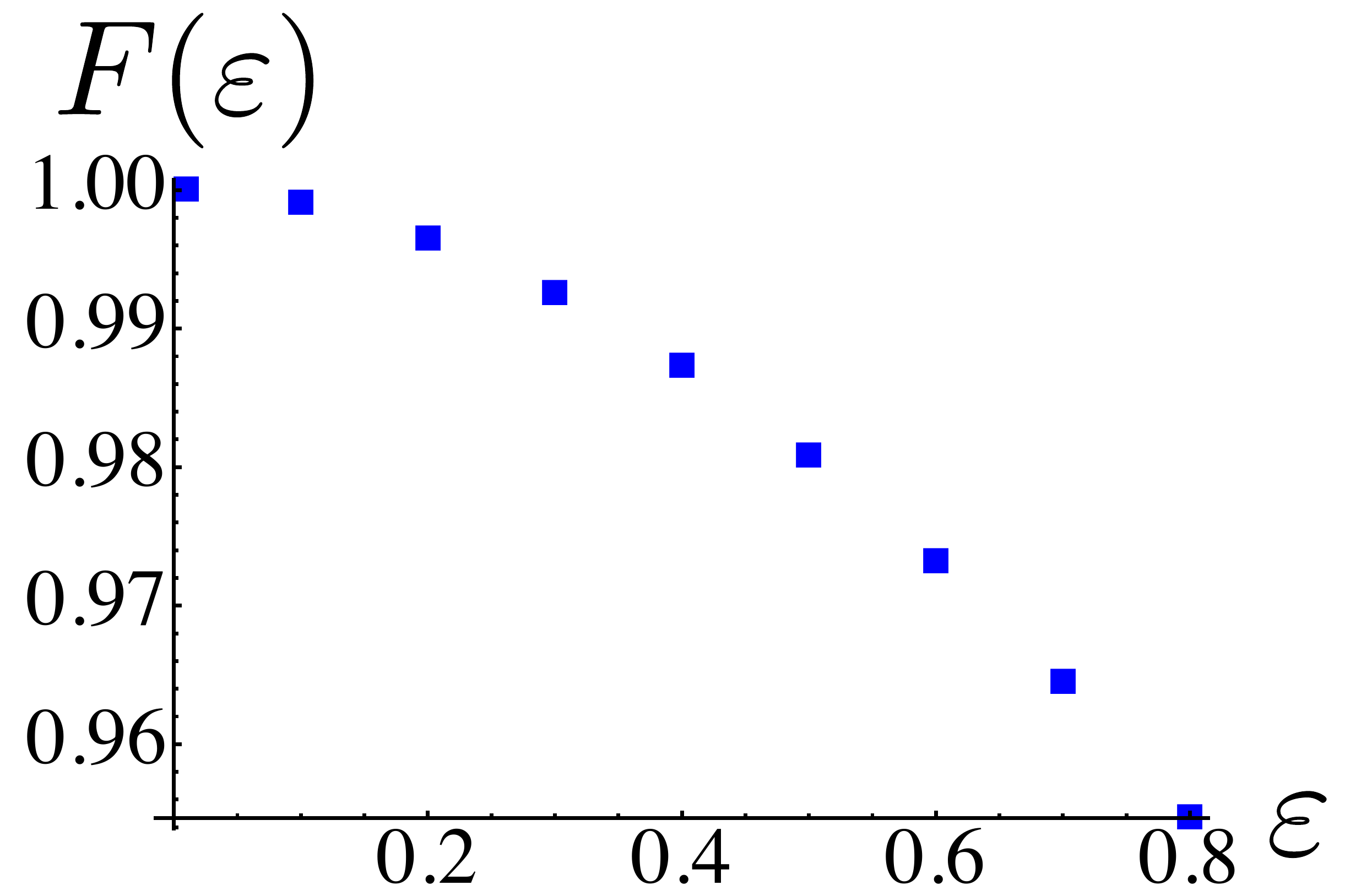}
\caption{(Color online) State fidelity $F(\varepsilon)$ between the
ground state of the undriven Duffing oscillator achieved through a
perturbative approach $\ket{\psi_{0}}$ and that estimated numerically by
diagonalising the Hamiltonian model in Eq.~\eqref{modello} using the
first 51 number states.} \label{fidelity}
\end{figure}

In order to determine the
range of values of $\varepsilon$ within which the form of $\ket{\psi_0}$
given in Eq.~\eqref{eq:GS} holds, we have calculated numerically the GS
$\ket{\psi_{num}}$ of the Hamiltonian $\hat H_{sys}$ in
Eq.~\eqref{modello} using a truncated Hilbert space consisting of the
first 51 number states and evaluated the state fidelity
$F(\varepsilon)=|\langle\psi_0|\psi_{num}\rangle|^2$, whose behavior
against the nonlinearity parameter $\varepsilon$ is shown in
Fig.~\ref{fidelity}. Fidelity remains above $95\%$ for
$\varepsilon\in[0,0.8]$. All the results reported in the remainder of
this paper have been gathered using this range of values. As we show
now, both the information on the modified potential energy and the GS
wave function are key for the analysis at the core of this paper. 
\par
We now pass to the introduction of the quantitative tools that we plan
to adopt in order to gather insight into the relation between
nonlinearity and the onset of non classicality in the Duffing oscillator
at hand. We would like to stress that the figures of merit that we introduce here go beyond the mere quantification of nonlinearity as given by the Hamiltonian parameters (e.g., as given by $\epsilon$ itself). This has the twofold advantage of allowing to encompass situations in which more than one Hamiltonian parameter is considered (see next Section) and of removing the dependence on the detailed form of the non-linear potential.

We first consider a measure of nonlinearity based on the
features of the GS~\cite{paris14} and built by comparing
$|\psi_0\rangle$ and its unperturbed counterpart $|0\rangle$.
Quantitatively, we determine the {\it distance} between the two GSs
using the Bures measure: Given a perturbing nonlinear potential $\hat
V(\hat x)$, we define the nonlinearity measure
$\eta_{\text{B}}\left[V\right]$ as the suitably normalized Bures
distance $D_\sb$ between the GS of the oscillator under consideration
and that of the corresponding harmonic one. In our case, we have 
\begin{equation}
\eta_\sb\left[\hbox{V}\right]=\frac1{\sqrt{2}}\,D_\sb \left[| \psi_0\rangle,
|0\rangle
\right]=\sqrt{1-|{}\langle 0
 \left|\psi_0\rangle\right| },
\label{eq:etaA}
\end{equation} 
where we have used the fact that the two states under scrutiny are pure. 
As it is apparent from its very definition, this quantifier depends
crucially upon the choice of a corresponding {\em reference harmonic
potential}. Such a dependence can be overcome by considering a second
way of quantifying nonlinearity: Given a potential $\hat {V}(\hat x)$
with associated GS $\ket{0}_V$, we define the measure of nonlinearity
$\eta_\sg \left[\hbox{V}\right]$ 
\begin{equation}
\eta_\sg \left[\hbox{V}\right]=\delta_\sg
\left[| 0\rangle \langle 0| _\sv\right],
\label{etaNG}
\end{equation}
where $\delta_\sg[\varrho ]$ is the degree of non-Gaussianity introduced
in Ref.~\cite{ng2,ng3}. This definition is intuitive: as commented
earlier, a nonlinear potential would induce deviations from Gaussianity,
which can in turn be used to quantify the strength of the nonlinear
process itself. The degree of non-Gaussianity is built on the quantum
relative entropy of the state and a {\it reference Gaussian state}.  As
the GSs under scrutiny are always pure, we have 
\begin{equation}
\eta_{\text{NG}}\left[\hbox{V}\right]
=S[\tau]=h\left(\sqrt{\hbox{det}[{\boldsymbol
\sigma}_\tau]}\right) \label{eq:etaApure}
\end{equation}
with $\tau$ the reference Gaussian state, $S[\tau]$ its Von 
Neumann entropy, $\boldsymbol{\sigma}_\tau$ its covariance matrix, and
$h(x)=(x+\frac{1}{2})\ln(x+\frac{1}{2})-(x-\frac{1}{2})\ln(x-\frac{1}{2})$.
The crucial point here is that the definition of $\eta_{\text{NG}}$
requires the determination of a reference Gaussian state for the GS of
$V(x)$ rather than a {reference harmonic potential} for $V(x)$ itself.
Both measures are zero for a harmonic potential, whereas they may lead
to different definitions of maximally nonlinear processes. 
\begin{figure}[t!] 
\psfrag{a}{\large$\varepsilon$}
\psfrag{b}{\large$\eta\left[\text{V}\right]$}
\includegraphics[width=0.495\columnwidth]{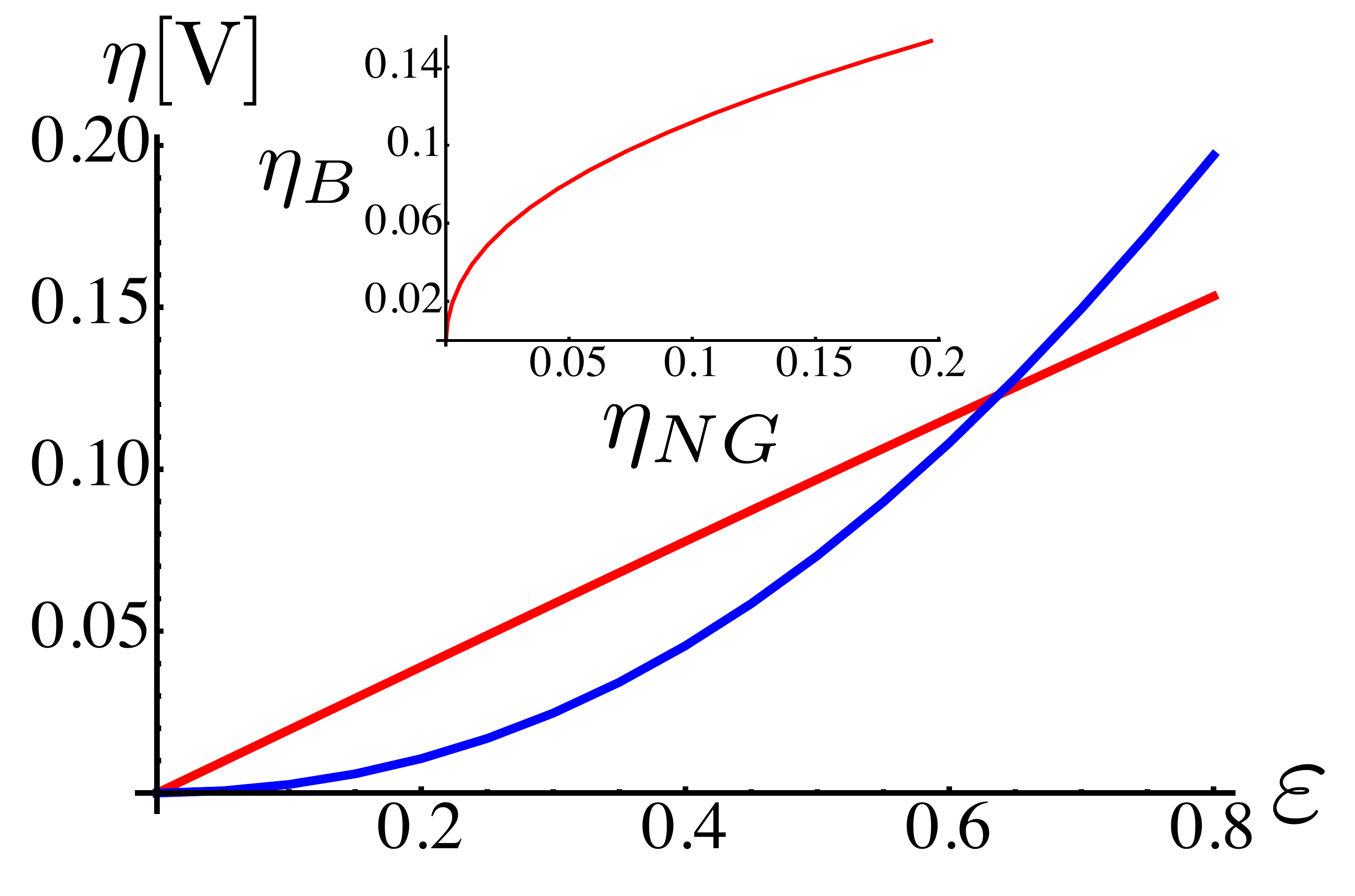}
\includegraphics[width=0.495\columnwidth]{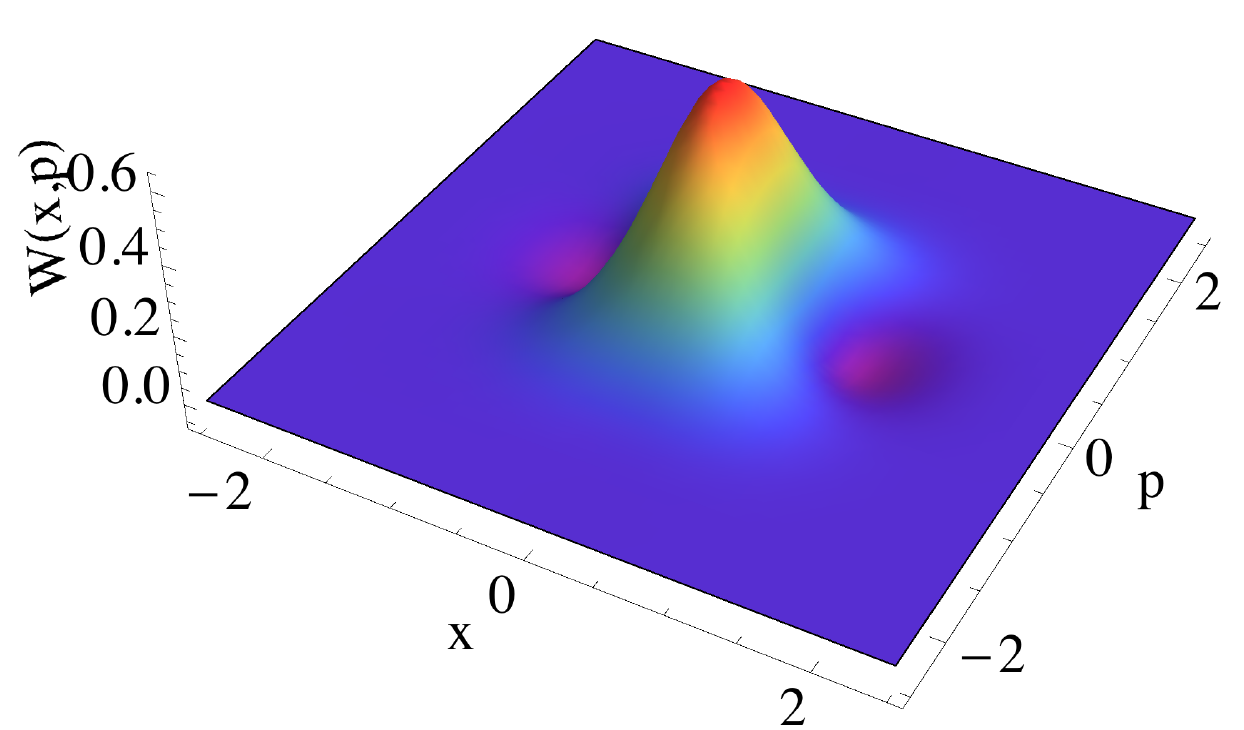}
\caption{(Color online) Left: the nonlinearity measures $\eta_\sb
\left[\text{V}\right]$ (red line) and $\eta_\sg \left[\text{V}\right]$
(blue line) for the undriven oscillator as a function of $\varepsilon$.
The inset is a parametric plot of $\eta_\sb$ as a function of
$\eta_\sg$, showing that the two measures are monotone functions of each
other.  The right panel shows the Wigner function of the GS for
$\varepsilon=0.1$
}
\label{f:F1bis}
\end{figure}
\par
While $0\le\eta_{B}[\hbox{V}]\le1$, the upper bound being reached if and
only if the external potential $\hat V(\hat x)$ gives rise to a GS
orthogonal to that of the corresponding harmonic case,
$\eta_{\text{NG}}$ is unbound from above, which complicates the
quantitative comparison between the two figures of merit. A suitable
rescaling can be obtained at any fixed value of energy upon normalizing
$\eta_{\text{NG}}$ to the degree of non-Gaussianity of the states that
achieve maximal non-Gaussian character at that value of the energy. 
This class includes number states and some specific superposition 
of them (see \cite{ng3} for details). The maximum of this rescaled 
quantity is thus achieved for a potential having a GS equal to a
number state ($n \neq 0$) of the harmonic oscillator or to some specific 
superpositions of them.
\par
One would expect that the nonlinearity increases with $\varepsilon$.
Indeed, as shown in Fig.~\ref{f:F1bis}, this intuitive behaviour is
captured by both measures, $\eta_{\text{NG}}\left[\hbox{V}\right]$ and
$\eta_{\text{B}}\left[\hbox{V}\right]$, which grow continuously and
smoothly with $\varepsilon$. The two measure are also linked by a
monotonic relationship. This is demonstrated in the inset of
Fig.~\ref{f:F1bis}, where we show a parametric plot of
$\eta_\sb\left[\hbox{V}\right]$ against $\eta_\sg\left[\hbox{V}\right]$,
the curvilinear abscissa in such plot being embodied by $\varepsilon$.
Numerically, for $\varepsilon\in[0,0.8]$, the relation between the two
measures of nonlinearity is very well approximated by the function
\begin{equation}
\eta_B[\hbox{V}]=a+b \sqrt{\eta_{NG}[\hbox{V}]}
\end{equation}
with $a\simeq2.7\times10^{-3}$ and $b\simeq0.34$.  
\par
A broadly used indicator of nonclassicality in the state of an
oscillator is provided by the volume occupied by its associated Wigner
function in the negative region of the phase space~\cite{ana}. The
Wigner function of the state $\rho$ of a single oscillator system is
defined as 
\begin{equation}
W_\rho(\alpha)=\frac{1}{\pi}\int 
e^{\alpha\xi^{*}-\alpha^{*}\xi}\chi_\rho(\xi)  d^2\xi,
\label{eq:Wig1}
\end{equation}
where $\chi_\rho(\xi)=\Tr[\rho e^{\xi\hat a^{\dagger}-\xi^{*}\hat a}]$
is the Weyl characteristic function and $\xi,\alpha\in{\mathbb C}$.
Unlike a true probability distribution, the
Wigner function can take on negative values \cite{wig,wig84,cah}, 
which is a
striking signature of nonclassicality. In the right panel of Fig. 
\ref{f:F1bis} we show the
Wigner function of the GS of the Duffing oscillator, showing the
presence of regions of negativity that signal the nonclassical nature of
the state of the system. In order to quantify such nonclassicality we
use the measure 
\begin{equation}
\label{nonClass}
\nu(\rho)=\frac{\eta_{\rho}}{1+\eta_\rho}
\end{equation} 
with 
$\eta(\rho)=\int_{-\infty}^{\infty}|W_\rho(\alpha)|d^2\alpha-1$ 
the negative volume of the Wigner function.
The quantity $\eta(\rho)$, which is per se sufficient to characterize
phase-space nonclassicality, has been employed to study the
quantum-to-classical transition in both linear and nonlinear
oscillators~\cite{katz, kleckner}, as well as to characterize the
performance of conditional schemes for the preparation of nonclassical
states of massive oscillators~\cite{Rogers,paternostroPRL}. Here we
consider its rescaled version according to Eq.~(\ref{nonClass}), which
provides a number $\nu(\rho)\in[0,1]$ that is thus amenable to a
quantitative comparison with the proposed measures of nonlinearity.
\par
\begin{figure}[h!] 
\includegraphics[width=0.99\columnwidth]{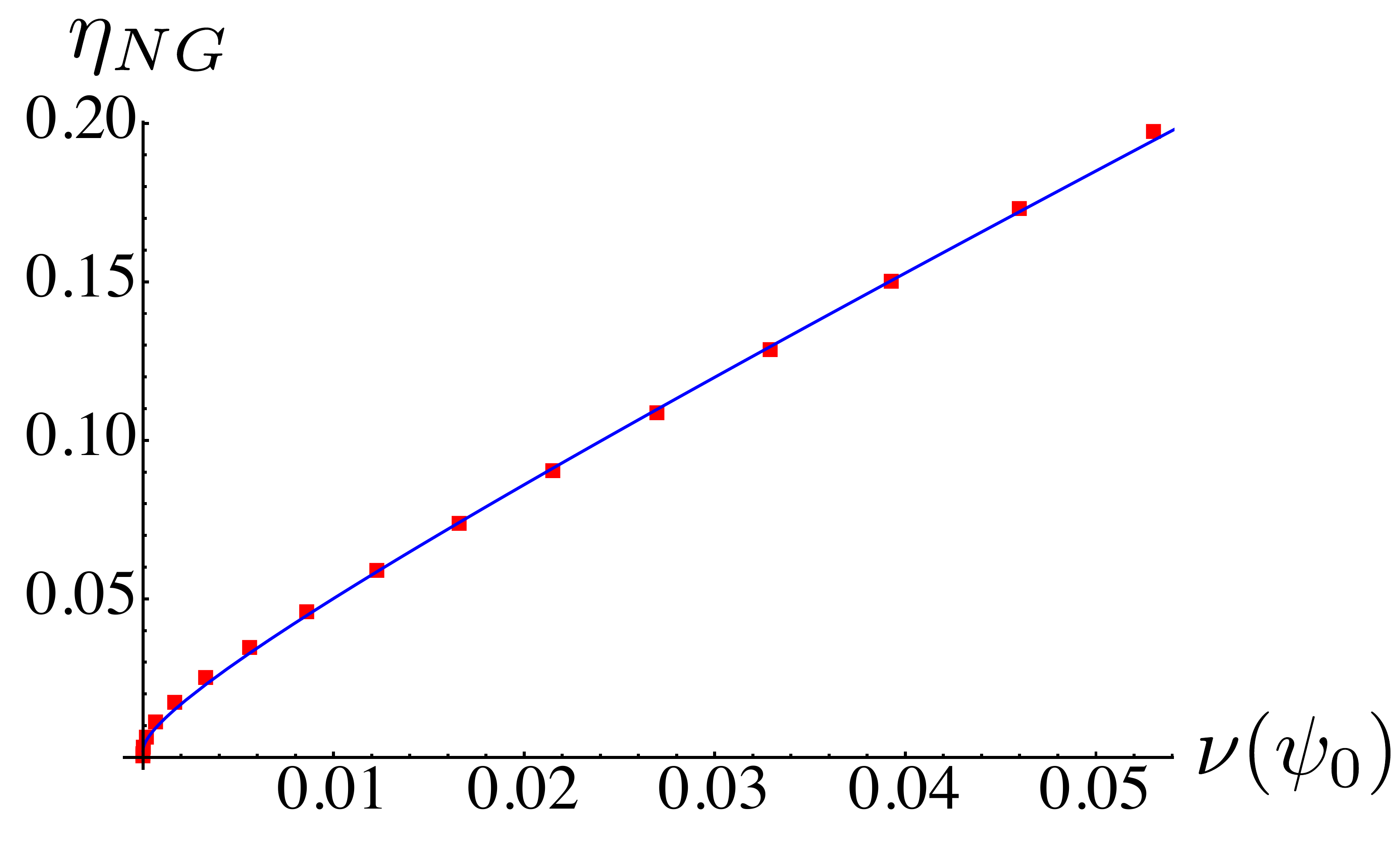}
\caption{ (Color online) We plot the  the nonlinearity
measure $\eta_\sg \left[\text{V}\right]$ against the normalized measure of $\nu(\psi_{0})$
nonclassicality based on the negative parts of the Wigner-function for
the GS of the undriven Duffing resonator. We show that the two measures
are monotone functions of each other. The squares show the values of $\eta_{NG}$ at set 
nonclassicality of the ground state of the undriven Duffing oscillator. The solid line represents the fit given by 
Eq.~(\ref{fit}). After an initial trait where $\eta_{NG}$ grows as
$\sqrt{\nu(\psi_0)}$, the relation between the two figures of merit
becomes approximately linear, showing the direct connection between
nonlinearity and nonclassical character of the state of the oscillator.} 
\label{f:F2} 
\end{figure}

Making use of Eq. ~(\ref{eq:Wig1}) for the Wigner function of the GS of
the Duffing resonator, we can evaluate the measure of nonclassicality
through a numerical integration.  In Fig.~\ref{f:F2} we show the plot of
the NG-based nonlinearity measure $\eta_\sg $ against the normalized
measure of nonclassicality $\nu(\psi_{0})$. 
 A numerical nonlinear fit gives the functional relation
\begin{equation}
\eta_{NG}=0.002+0.207 \sqrt{\nu(\psi_0)}+2.731\nu(\psi_0),
\label{fit}
\end{equation}
showing that after an initial trait, the link between the two quantities
becomes linear, ensuring the proportionality of the two figures of merit
under scrutiny. 
\par
\section{Nonlinearity of a Duffing oscillator: driven case}
\label{s:DRIVEN}
\begin{figure}[h!] 
\includegraphics[width=0.53\columnwidth]{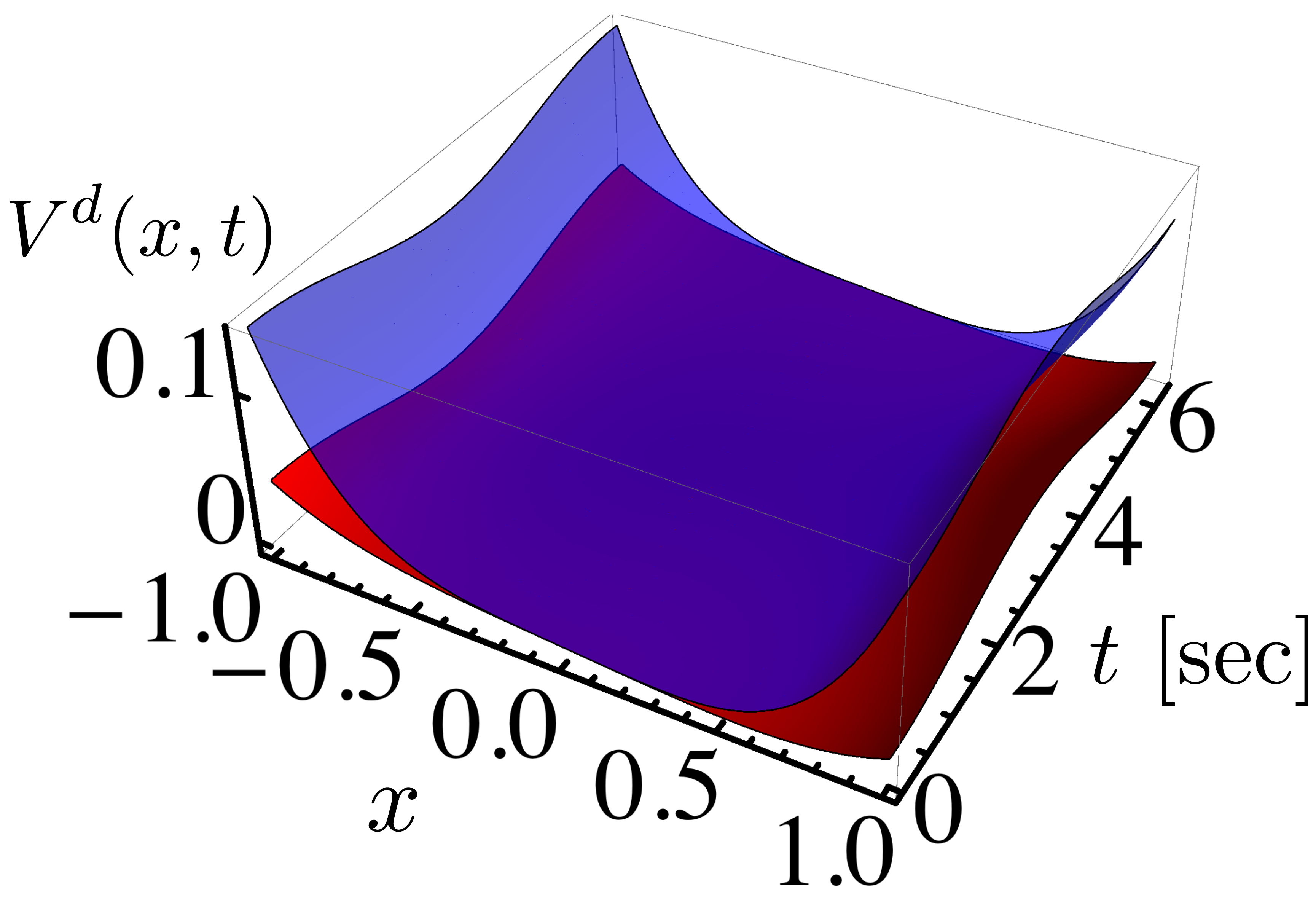}
\includegraphics[width=0.45\columnwidth]{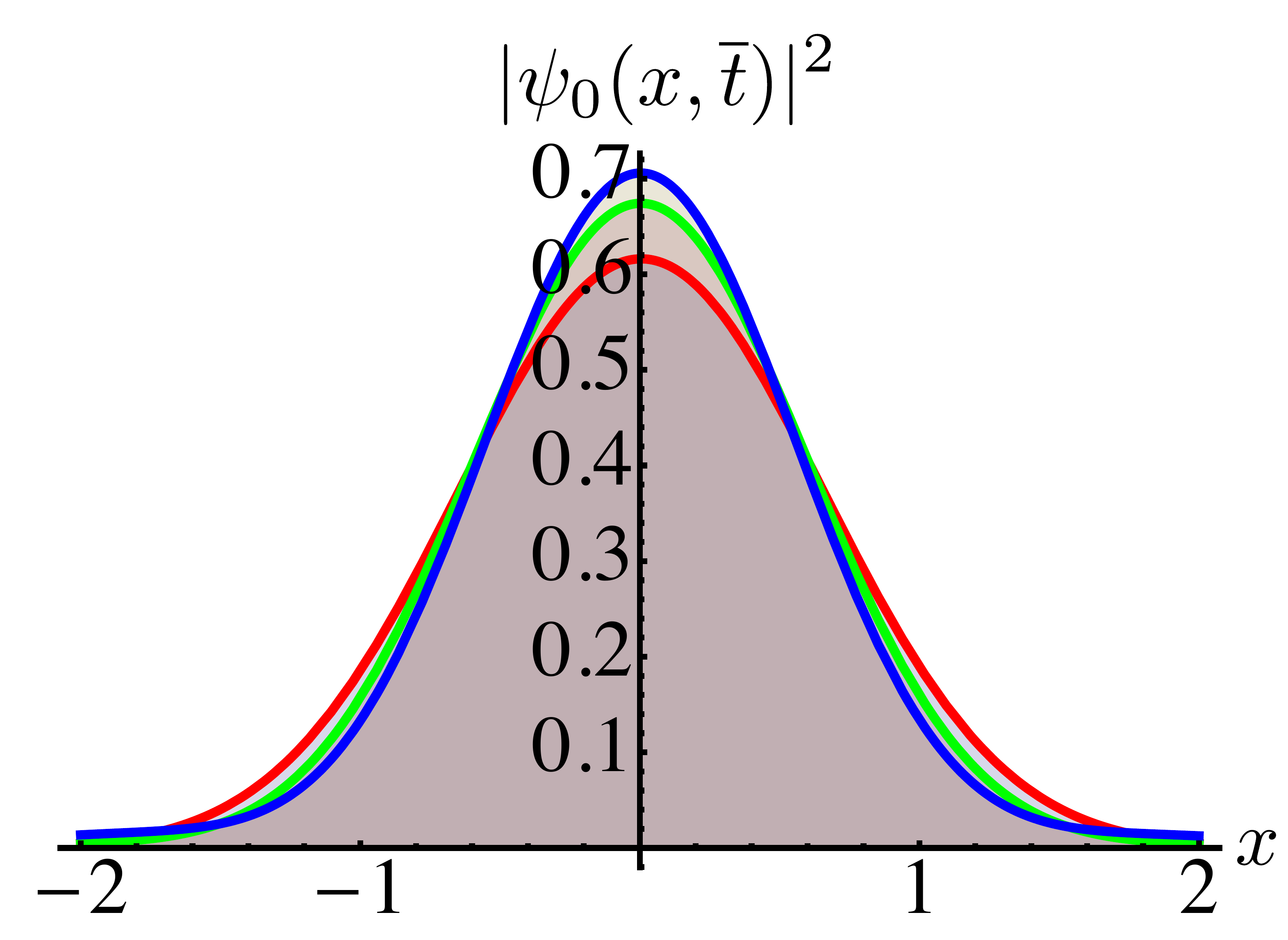}
\caption{ (Color online) In the left panel we show $V^d(x,t)$ 
against time and position for ${\cal F}=0.015$, $\varepsilon=0.1$ 
(red, opaque surface), and $\varepsilon=0.5$ (blue, transparent surface). 
In the right panel we show the corresponding GS probability densities, 
$|\psi_0(x,\overline{t}=1)|^{2}$ for the same set of parameters.}
\label{f:F5}
\end{figure}

We shall now pass to the analysis of a driven Duffing resonator, which
is an example of  nonlinear resonator often encountered in relevant
experimental situations~\cite{craig,postma,kozinsky,aldridge}.  The
dimensionless Hamiltonian of the system is thus
Eq.~\eqref{HamiltonianModel} with the explicit inclusion of the
time-dependent driving term $-\hat x {\cal F}\cos\omega t$. In what
follows, we choose a working point well within the region of bistability
of the oscillator (which is ensured for ${\cal F}\in[0.015,0.06]$ and
$\omega\in[1.016,1.02]$). The form of the driving potential ${\hat
V}^d(\hat x,t)=\frac{1}{4}\varepsilon \hat x^4-\hat x {\cal F}\cos\omega
t$ is illustrated in Fig.~\ref{f:F5}, showing the deformation
induced by the nonlinear and time-dependent part of the perturbation.
In order to evaluate the form of the GS associated with the full model, 
we shall resort to the use of time-dependent perturbation theory. 
We decompose the state of the system at $t=0$, when the perturbative 
potential is off, as 
\begin{equation} 
|\psi(0)\rangle=\sum_{n}c_n{(0)}|n\rangle~~~(c_n(t)\in{\mathbb C})
\end{equation} 
and aim at finding a perturbative expansion $c^{(q)}_n(t>0)
\in{\mathbb C}$ at order $q\in{\mathbb Z}$ in the perturbation, 
so that the state of the system at the corresponding order becomes 
\begin{equation} 
|\psi^{(q)}(t)\rangle\simeq\sum_{n}c^{(q)}_n(t)
e^{iE_{n}t/\hbar}|n\rangle,
\end{equation} 
where $E_n$ is the eigenvalue of the unperturbed Hamiltonian corresponding to the eigenstates $\ket{n}$.
The first-order correction in both ${\cal F}$ and ${\varepsilon}$, which
will be the highest order of the perturbative expansion that we will
consider here, is given by~\cite{sak} 
\begin{equation}
c^{(1)}_{n}(t)=-{i}\int_{0}^{t}
e^{i\omega_{n\ell}t^{\prime}}V^d_{nl}(t^{\prime})dt^{\prime},
\end{equation}
with $c^{(0)}_{n}(t)=\delta_{nl}$, $V^d_{nl}(t)=\langle n|\hat V^d(\hat x,t)|l\rangle$ and $e^{i(E_{n}-E_{l})t} =e^{i\omega_{nl}t}$.
The explicit calculation of $|\psi(t)\rangle$ up to 
the stated order of approximation and for $|\psi(0)\rangle=|{0}\rangle$ 
leads us to the GS wavefunction
\begin{equation}
\label{eq:GSD}
\begin{aligned}
\psi_0&(x,t)={\cal M}\left\{\bigg(1
-\frac{3it\varepsilon}{16}\bigg)e^{-it/2}\psi_0(x)\right.\\
+&{\cal F}\,\frac{(1-e^{it}[\cos\omega t
-i\omega\sin\omega t])}{\sqrt{2}(\omega^2-1)}e^{-3it/2}\psi_{1}(x)\notag\\
+&\frac{\varepsilon}{16}\!\Bigg[{3\sqrt{2}(e^{-5it/2}
-e^{-it/2})}\psi_{2}(x)
\notag \\
&\qquad{+} \frac{(e^{-\frac{9}{2}it}-e^{
-\frac{i}{2}t})}{4}\psi_{4}(x)\Bigg]\Bigg\}
\end{aligned}
\end{equation}
being ${\cal M}$ the normalisation factor 
and $\psi_n(x)\equiv\langle x|n\rangle$ the wave 
function of state $| n\rangle$. 
A plot of the spatial 
distribution $|\psi_0(x,\overline{t})|^2$ at a set 
time $\overline{t}$ is shown in the right panel Fig.~\ref{f:F5}. 
\begin{figure}[h!] 
\includegraphics[width=0.495\columnwidth]{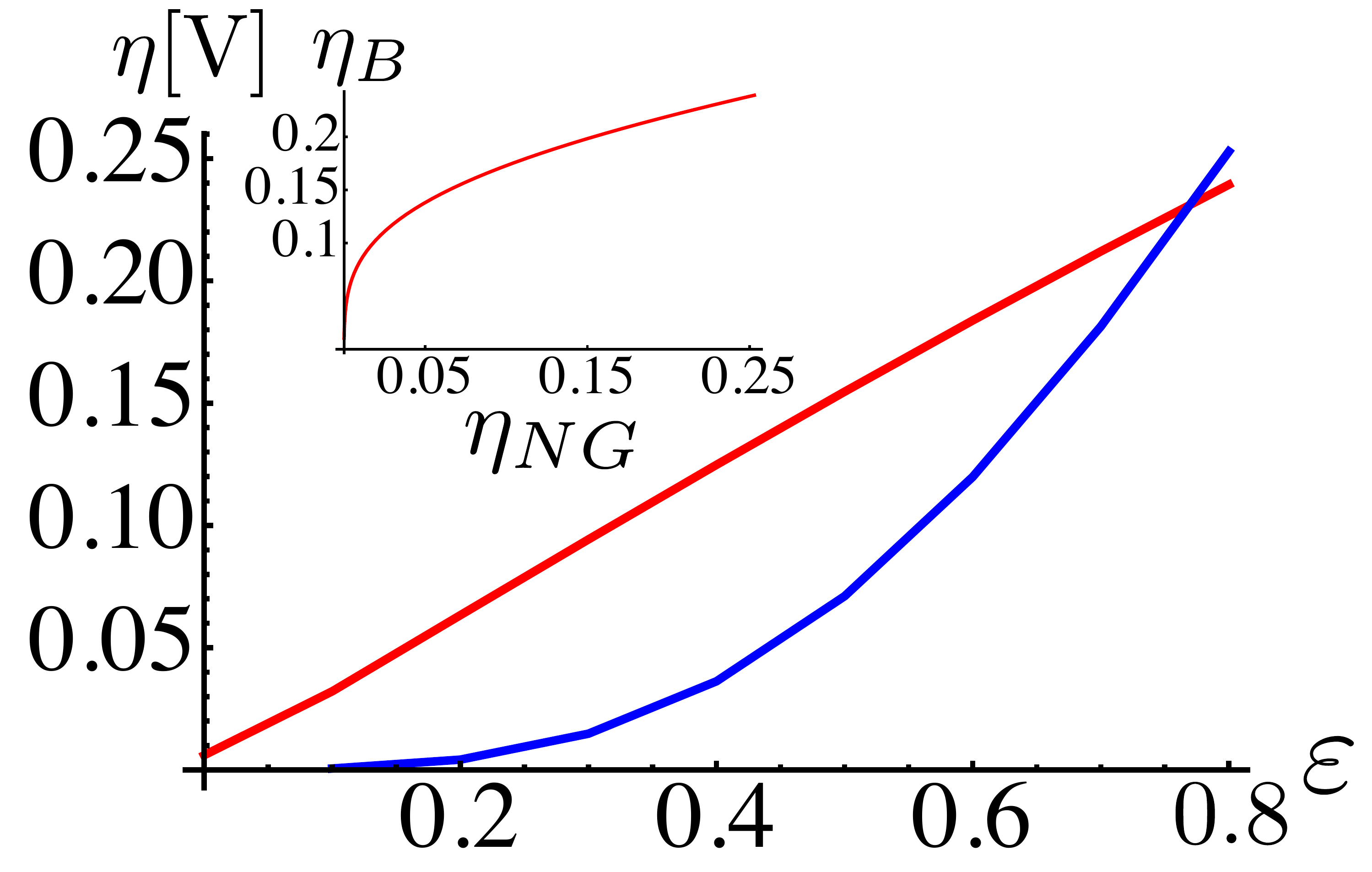}
\includegraphics[width=0.495\columnwidth]{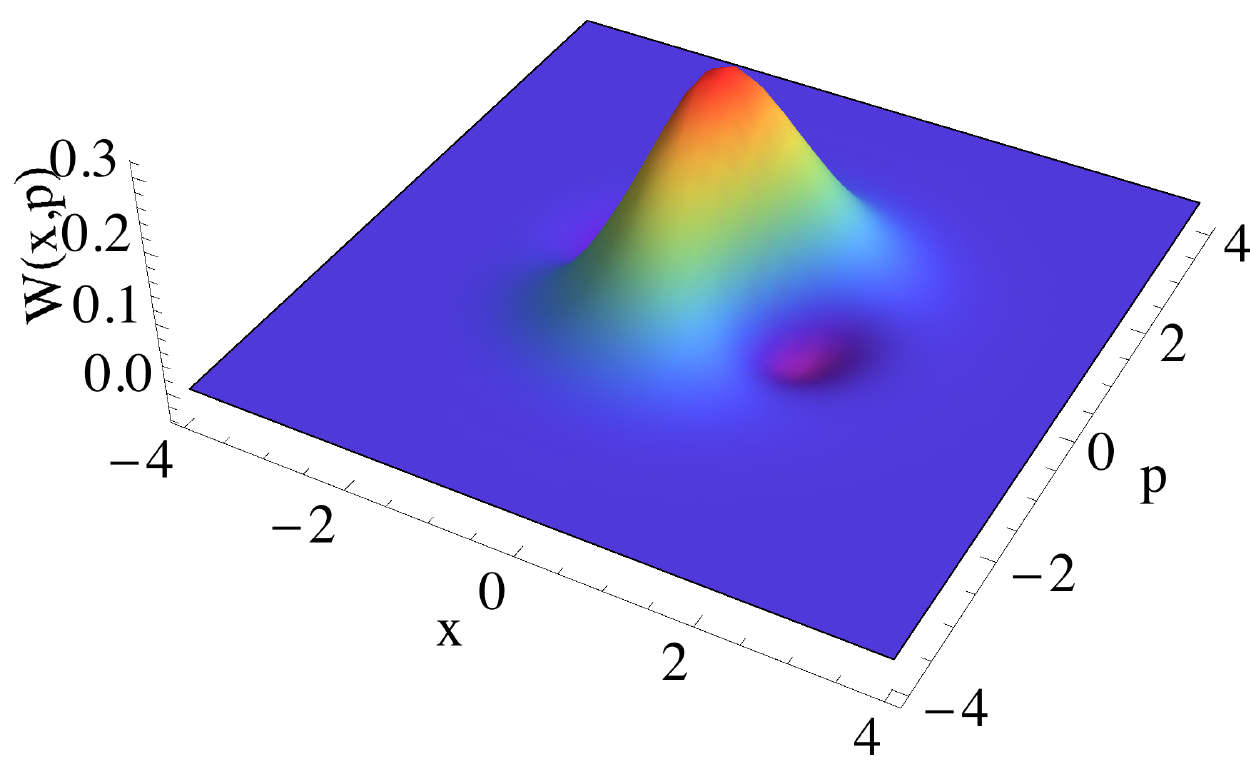}
\caption{(Color online) Left: the nonlinearity measures
$\eta_\sb \left[\text{V}\right]$ (red line) and $\eta_\sg 
\left[\text{V}\right]$ (blue line) for the driven oscillator 
as a function of $\varepsilon$ and for the choice of parameters ${\cal
F}=0.015$, $t=1$ and $\omega=1.018$ which ensures that  the resonator is
in the bistability region. The inset is a parametric plot of $\eta_\sb$
as a function of $\eta_\sg$, showing that the two measures are monotone
functions of each other.  The right panel shows the Wigner function for
the same set of parameters and $\varepsilon=0.1$. }
\label{f:F5bis}
\end{figure}
\par
\begin{figure}[h!] 
\includegraphics[width=0.99\columnwidth]{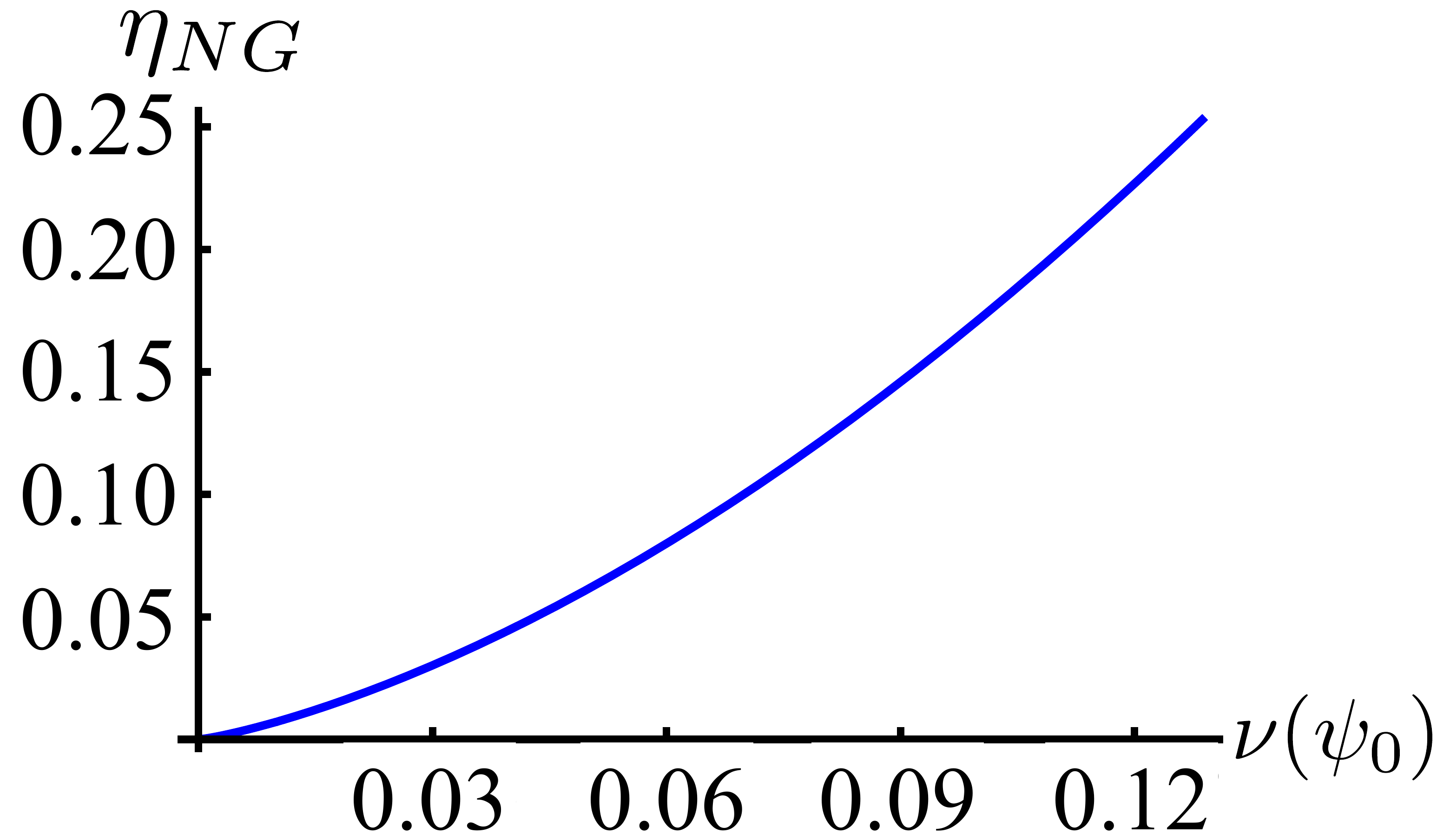}
\caption{(Color online) 
A parametric plot of the normalized measure of $\nu(\psi_{0})$
nonclassicality based on the negative parts of the Wigner-function for
the GS of the driven Duffing resonator as a function of the nonlinearity
measure $\eta_\sg \left[\text{V}\right]$ (parameters $\varepsilon=0.01$,
$t=1$, $\omega=1.018$, and ${\cal F}=0.015$) appears to be roughly linear and monotonically
increasing.} \label{f:F6aa}
\end{figure}

Looking at the potential, one would expect that the 
nonlinearity increases with $\varepsilon$ at any fixed 
values of the given parameters. Indeed, this intuitive 
behavior is captured by both measures, $\eta_\sg \left[
\text{V}\right]$ and $\eta_\sb \left[\text{V}\right]$, 
as they grow continuously. The two measures are monotonic 
functions of each other, as illustrated in the left panel of 
Fig. \ref{f:F5bis}. In the right panel we also show  
the Wigner function 
for $t=1$, which shows negative regions and thus the 
signature of nonclassicality.
When assessed against the chosen indicator of nonclassicality,
$\eta_{NG}$ is again found to be in direct correspondence with the
volume occupied by the Wigner function corresponding to
$|\psi_0(t)\rangle$ [cf. Fig.~\ref{f:F6aa}]. 
\par
\section{Conclusions}
\label{conc}
In this paper, upon using purpose-tailored quantitative 
figures of merit, we have addressed in some details the relation 
between nonlinearity and nonclassicality in a class of
nonlinear oscillators that is relevant in many contexts, including
experimental solid state physics. By approximating the form of the
ground state of the oscillator through a perturbative approach (either
stationary or time-dependent) we have been able to demonstrate that the
negativity of Wigner function, which is a well-acquired measure of
nonclassicality in continuous variable systems, is in monotonic relation
with recently proposed measures of nonlinearity, therefore reinforcing
the idea of nonlinearity as a catalyst of quantumness. Although our
conclusions have been gathered by addressing the specific example of 
a Duffing oscillator, our work paves the way to interesting 
extensions, primarily concerned with the application of our tools 
to other forms of nonlinearities~\cite{tocome}. 
\par
A second interesting direction of investigation would deal with the
inclusion of environmental effects, and with the possibility to shield the
degree of nonclassicality enforced in the state of a quantum oscillator
through a suitable degree of nonlinearity. This might entail an
interesting way of protecting quantumness, stemming from the direct,
non-demanding control of the Hamiltonian of the oscillator. In fact,
while the harmonic assumption is an approximation valid within many
contexts (from nano-mechanical oscillators to ultracold atomic systems
in external potential), switching to explicitly non-harmonic situations
is typically straightforward by the means of a strong driving. This is generally more economic than time-gated external pulses (required in
dome of the techniques devised so far for the protection of quantum
features) or the control of the properties of the environment, which is
typically of not easy access.  Work along these lines is in progress and
results will be presented elsewhere.
\section*{Acknowledgments}
This work was supported by MIUR (FIRB ``LiCHIS'' - RBFR10YQ3H), the UK
EPSRC (EP/G004579/1), and the John Templeton Foundation (grant ID
43467). B.T. was supported by the TRIL Programme of ICTP, and 
acknowledges hospitality at the Centre for Theoretical Atomic,
Molecular, and Optical Physics, School of Mathematics and Physics,
Queen's University Belfast, where the initial plans of this project were
conceived.

\end{document}